\documentclass[prc,12pt,english]{revtex4} 
\usepackage{babel}
\usepackage{amsmath} 
\usepackage{graphics}
\usepackage{pdfpages} 

\newcommand\ba{\begin{eqnarray}}
\newcommand\ea{\end{eqnarray}} 
\newcommand\be{\begin{equation}}
\newcommand\ee{\end{equation}} 
\newcommand\alb{\begin{align}}
\newcommand\ale{\end{align}}

\usepackage{amsmath} % Enhanced mathematics \usepackage{amssymb}
\begin{document}

\title{Compared analysis of time-like hyperon form factors}

\author{Andrea Bianconi} \email{andrea.bianconi@unibs.it}
\affiliation{\it Dipartimento di Ingegneria
dell\!~$^\prime$Informazione, Universit\`a degli Studi di Brescia and
Istituto Nazionale di Fisica Nucleare, Gruppo Collegato di Brescia,
I-25133, Brescia, Italy}

\author{E.~Tomasi-Gustafsson} \email{egle.tomasi@cea.fr}
\affiliation{\it DPhN, IRFU, CEA, Universit\'e Paris-Saclay, 91191
Gif-sur-Yvette Cedex, France}

\begin{abstract} 
Data on hyperon form factors recently obtained  from
annihilation reactions are compared each other, and to proton and neutron form factors, 
in terms of two kinematical variables: the transferred momentum square,
$q^2$, and the modulus $P$ of the relative momentum between the outgoing baryons. They are critically discussed in terms of possible correlated structures. The
present status of the time-like form factor data for baryons is
described and suggestions are given on the reactions and the kinematical
range where data are desirable in order to clarify the  arisen
questions.

\end{abstract} 
\date{\today} 
\maketitle

\section{Introduction }
\subsection{General background }

The internal properties of composite particles are conveniently
expressed in terms of form factors (FFs) (for a review see
\cite{Pacetti:2015iqa,Denig:2012by}). FFs are measured in elastic $eh\to
eh$ scattering and in annihilation reactions as $e^+e^-\leftrightarrow h
\bar  h $,  where $h$ is a hadron. Form factors are Lorentz scalar functions of the only one
variable $q^2$, being $q$ the four-momentum of the virtual photon which
mediates in the $t$ and $s$-channel, respectively, the scattering and
the annihilation processes in the Born approximation. The scattering reaction covers the
region of space-like momenta, $q^2 < 0$, whereas the annihilation
reactions give information on the time-like region of transferred
momenta, $q^2 > 0$. 

This work focuses on FFs data in the time-like region,
from electron positron colliders. Although data on the proton and
neutron time-like electromagnetic FFs (TLFF) have a long
history, the precision of recent experiments has recently opened a
debate on aspects of the TLFF that cannot be simply deduced or guessed
by extrapolating space-like properties.

Since the  FENICE experiment at Frascati in 1994~\cite{Antonelli:1993vz}, 
new precise measurements from modern colliders added precious information in the time-like region, 
particularly on neutrons and hyperons. The high luminosity
obtained at the present $e^+e^-$ colliders opens the opportunity to apply the quasi real electron
method~\cite{Baier:1980kx}, also called initial state radiation (ISR)
technique, to experiments like
BABAR and BESIII. It allows to extract $e^+e^-$-annihilated cross section
values from the differential cross section data of processes where a
real hard photon is emitted from one of the lepton beams and is
detected, allowing the selection of the
four-momentum transferred to the final hadron pair, as for a beam energy-scan. 
The large centre of mass energy of modern colliders brings the advantage to obtain, by means
of the ISR technique, data in a wide region of $q^2$ and with a high
level of accuracy, despite the reduction of the cross section by a
factor of $\alpha_{\rm EM}\simeq 1/137$, the fine structure constant,
which is indeed compensated by the high luminosity.

BABAR gave first very precise measurements on proton FF in a wide
region, up to 40 points for $4.25 \le q^2 \le 36 $ GeV$^2$ \cite{Lees:2013xe,Lees:2013uta}. These data show evidence of specific structures, that become regular when
plotted as a function of $P$,  the modulus of relative three-momentum of
the proton-antiproton pair in the final state, as discovered  in a
series of articles \cite{Bianconi:2015owa,Bianconi:2016bss,Bianconi:2018unb}. Repeating the
same analysis, these oscillations were confirmed on the proton by BESIII
both with the ISR method \cite{BESIII:2015axk,BESIII:2021rqk} and with
the energy scan method \cite{BESIII:2019hdp}, and more recently on the
neutron FF data, described by oscillations differing from the proton
ones by a phase \cite{BESIII:2021tbq}.

Hyperon TLFFs were also measured, although with less precision with
respect to the nucleon FFs. The BESIII
experiment~\cite{BESIII:2017hyw} measured the
$e^+e^-\to\Lambda\bar\Lambda$ cross section at four energies, namely:
$\sqrt{s} =2.2324$, 2.400, 2.800 and 3.080 GeV, lying in the same range
already investigated by the BABAR experiment, but with better precision.
Recently, high energy cross section measurements were given in
Ref.~\cite{BESIII:2021ccp} with the aim to find evidence of the
$\psi(3770)$ vector meson. The BESIII experiment~\cite{BESIII:2020uqk}
measured at six energy points the $\Sigma^+\bar\Sigma^-$ and at three
points the $\Sigma^-\bar\Sigma^+$ cross
sections~\cite{BESIII:2020uqk,BESIII:2021rkn} by extracting the
corresponding TLFFs, while the BABAR experiment measured the
$\Sigma^0\bar\Sigma^0$ as well as the $\Lambda\bar\Sigma^0+$c.c. cross
sections, giving their TLFFs and transition TLFFs~\cite{Aubert:2007uf},
respectively. The CLEO experiment~\cite{Dobbs:2017hyd} also measured
cross sections and extracted TLFFs of several hyperons in 
the kinematical region around the mass of the $\psi(3770)$ vector meson.
These data will not be used in the present analysis, as they are less precise
than the most recent ones.

The BESIII experiment has also obtained values of the TLFFs of the $\Xi$
hyperons, namely: eight values for the charged
$\Xi^-$~\cite{BESIII:2020ktn} and ten for the neutral
$\Xi^0$~\cite{BESIII:2021aer}, all of them belonging to the energy
interval $[2.644,~3.080]$~GeV.

Note that, although the baryons $\Sigma^+$, $\Sigma^0$ and $\Sigma^-$ form
an isospin triplet, the $\Sigma^-$ is not the antiparticle of the
$\Sigma^+$, so there are two different TLFFs corresponding to the final
state $\Sigma^+\bar\Sigma^-$ and $\Sigma^-\bar\Sigma^+$. 

All along the paper, natural units are used, so that $\hbar =c=1$.

%%%%%%%%%%%%%%%%%%%%%%%%%%%%%%%%%%%%%%%%%%%%%%%%%
\subsection{Irregularities in the proton TLFFs}
\label{subsec:intro2}
Thanks to the high luminosity available at the colliders, only recently  experiments 
could collect enough statistics to make possible the extraction of the individual 
moduli of the electric and magnetic TLFFs. However, from the cross section, 
of the annihilation $e^+e^-\to h\bar h$, where $h$ 
is a spin-1/2 hadron, in the Born approximation one can always define the effective TLFF $F_h$ as 
\ba
F_{h}^2(q^2)\ \equiv\ A_E(q^2) |G_E(q^2)|^2 + A_M(q^2) |G_M(q^2)|^2
\label{Eq:Fp} 
\ea 
where $A_E$ and $A_M$ are kinematic coefficients,  $q$ 
is the four-momentum of the $h\bar h$ system, and hence $q^2$ 
is the squared value of the center of mass energy. The total cross section in this frame can be written as: 
\ba
\sigma(q^2)=K(q^2)\,F_h^2(q^2)\,,
\ea
where $K(q^2)$ is a known kinematic factor. 
An interesting aspect is the presence of small oscillations in the proton effective TLFF,
the systematic character of which has been highlighted in
Ref. \cite{Bianconi:2015owa},  after decomposing  $F_p$ as a sum of two terms: a regular "background" 
and an "oscillation" term, namely   
\ba
F_p(q^2)\ \equiv\ F_{\rm bkg}(q^2)\ + F_{osc}(P), 
\label{eq:osc0} 
\ea 
where  $P$  is the modulus 
of the relative three-momentum of the final $p\bar p$ pair (see later for the definition).

The background term describes a regular decrease not far from the known
$(1/q^2)^2$ quark counting rule \cite{Matveev:1973uz,Brodsky:1973kr}
while the modulation term has the form 
\ba 
F_{\rm osc}(P)\ =\ A\ e^{-BP}\cos(CP+D), 
\label{eq:osc1} 
\ea 
where $A$, $B$, $C$ and $D$ to be determined accordingly to the data. 
At the condition of a moderate
exponential decrease, this describes quasi-periodic oscillations. Several
background terms taken from the literature have been tested in Ref.
\cite{Bianconi:2015owa}, as a standard dipole, and the conclusions
did not depend on the choice. In all cases the fitted values
of $A$ and $B$ imply that the oscillation amplitude is
about 5-10 \% of the background in all the relevant range, $C$
corresponds to the oscillation period ${2\pi}/{C}\simeq 1.1$ GeV, 
and $D$ $\approx$ 0 indicates that there is an oscillation maximum at the $p\bar p$ production threshold, where $P=0$. 

Irregularities in a few-GeV hadronic state are not a surprise, but their
periodicity opened a question on the underlying mechanism. The modulation may be interpreted as a self-coherent pattern
(possibly of interference origin), but it may be a set of three
independent resonances, as suggested in Ref. \cite{Lorenz:2015pba}.
These two interpretations  are not necessarily antagonistic, if the
formation of a group of resonances is due to some simple and common
underlying mechanism. For example a toy model on the pion  TLFF predicts a periodic sequence of resonances \cite{deMelo:2003uk}.

By making Fourier transforms, the analysis carried out in
Ref.\ \cite{Bianconi:2015owa} associates the oscillation period to a spacetime
dimension of 1 fm, implying a non-scaling mechanism acting on
such a scale. As a consequence, one expects the oscillations
to disappear when $q^2$ becomes large enough to reach the expected
$\simeq (1/q^2)^2$ asymptotic behavior. 

A more theoretical analysis was carried out in Refs.
\cite{Bianconi:2018unb,Bianconi:2016bss}, showing that energy-periodic
oscillations are present when two almost identical, but slightly shifted
in time, formation pathways compete to reach the same final state. Since
the increasing space distance $r$ between the baryon and the antibaryon
in the final state is, in the average, proportional to the time, an {\it
interference in time}  corresponds to an {\it interference in space} and, hence, in the 
distance $r$. The hadron-antihadron relative three-momentum $\vec P$ 
is the conjugate variable of the distance  $\vec r$.

On the other side, the most recent BESIII data on proton TLFF have
included several values of the ratio $|R|=|G_E/G_M|$. Complemented by
BABAR \cite{Lees:2013xe,Lees:2013uta}  and Novosibirsk 
\cite{CMD-3:2018kql} data on  $F_p$  that span a region down to  threshold,  where  the
constraint $G_E(4M_p^2)=G_M(4M_p^2)$, holds (being $M_p$ the proton mass), a continuous
description  of $|G_E|$ and $|G_M|$ in a wide range  of $q^2$ was
given in the form of a physics driven fit
\cite{Tomasi-Gustafsson:2020vae}. $|G_E|$ and $|G_M|$ do not appear to
be periodic functions of $P$, in contrast with their  combination
$F_p$. Whether the periodicity of $F_p$ is there by mere chance or
not is still to be clarified. This leads to a more general question,
that is whether $F_p$ may have a meaning as an effective amplitude, or
must just be considered as the real and positive squared root of a
linear combination of $|G_E|^2$ and $|G_M|^2$ from Eq. (\ref{Eq:Fp}).

%%%%%%%%%%%%%%%%%%%%%%%%%%%%%%%%%%%%%%%%%%%%%%%%%%%%%
\subsection{Proton and neutron time-like form factors}
\label{subsec:intro3}
%%%%%%%%%%%%%%%%%%%%%%%%%%%%%%%%%%%%%%%%%%%%%%%%%%%
An analysis  of pre-BESIII data had shown the presence of near threshold
enhancements in neutron and neutral baryon TLFFs \cite{Baldini:2007qg}.
Later, BESIII data \cite{BESIII:2021tbq} on the neutron TLFF have shown oscillations as well. 
Adopting the technique of Ref.
\cite{Bianconi:2015owa}, the authors of this measurement have
interpolated the neutron FF data according to Eqs.
(\ref{eq:osc0},\ref{eq:osc1}), finding best-fit results with the same
period of the proton data (in the variable $P$) and a 120$^\circ$ phase shift. 
Two weak points of these results are (i) some
gaps in the data range leave ambiguity on the precise shape of the
neutron curve, (ii) by comparing the proton data fit of Ref.
\cite{BESIII:2021tbq} with those of Refs.
\cite{Bianconi:2016bss,Tomasi-Gustafsson:2020vae}, one finds that the
phase of the proton oscillating component depends on the choice of the background 
component when the expression of Eq.~\eqref{eq:osc0} is used. The oscillation period
is however reasonably background-independent.

A confirmation of this correlation by a more complete set of data would
support the presence of relevant channel mixing in the production mechanisms of 
$\gamma^* \rightarrow p\bar{p}$ and $\gamma^* \rightarrow n\bar{n}$. An analysis of the
constraint posed by the present data on this mixing has been carried
out in Ref. \cite{Cao:2021asd} under the assumption that the isoscalar, $I=0$, and isovector, $I=1$,  states are independently formed in the electromagnetic initial part of  the process and not later remixed. Although the confirmation of a channel mixing would be a relevant phenomenon in itself,
expressing $p\bar{p}$ and $n\bar{n}$ in terms of other states
does not explain the origin of the $p\bar{p}$ and $n\bar{n}$
irregularities. This means that either the isospin channels do
independently oscillate, or alternatively that they have an energy-dependent relative phase $\phi(q^2)$. In any case, the
origin of the oscillations or of $\phi(q^2)$ remains unknown.

Summarizing the state of the art with irregularities in the nucleon
TLFFs, it is evident that peculiar time-like
phenomena are present. It is not clear whether they derive from processes that may be similar, 
but independently affect the $p \bar{p}$ and $n\bar{n}$ channels, or are connected to some 
degree of channel mixing. It is also not clear whether we are seeing an overlap of 
independent processes (like three resonances) or the effects of a unique underlying 
mechanism (like two interfering amplitudes, 
or flux exchange between $p \bar{p}$ and $n\bar{n}$ final states). 

%%%%%%%%%%%%%%%%%%%%%%%%%%%%%%%%%
\subsection{Baryon form factors and aims of the present work}
\label{subsec:intro4}
%%%%%%%%%%%%%%%%%%%%%%%%%%%%%
In addition to nucleon TLFFs, several sets of data have been collected
on strange hyperon effective TLFFs. These data, at first sight, support the
possibility of additional non-monotonous and irregular behaviors. However, at present, 
their precision and distribution do not allow to shed light on the dynamical mechanisms 
underlying such behaviors. It is quite difficult reproduce 
these data in an unambiguous way, as evident in some of the proposed fits of
Ref.~\cite{Dai:2021yqr}. Apart from fitting difficulties, 
the question is what we may understand from these data on the underlying processes. 

The aim of the present work is to organize the hyperon data and find
correlations or common behaviors among them and with the nucleon TLFFs,
to gain further insight on the discussed problems.

Waiting for more precision and statistics in the measurements of the
years to come, we feel that it is better not to stick to very specific
models like the oscillations of Eq. (\ref{eq:osc1}). We will rather try
to identify correlations and common trends in the available data on
proton, neutron and several strange baryon TLFF, including data on the 
$\Lambda\bar\Sigma^0$ transition TLFF, 
which parametrizes the effective amplitude of the process 
$\gamma^*\to\Lambda\bar{\Sigma}^0+$ c.c.. 

The cases of the neutron and the proton, as well as their comparison, 
have been already discussed in Refs.~\cite{BESIII:2021tbq,Dai:2021yqr}. 
A dedicated analysis will be published in a fore-coming paper. 
We will only marginally discuss the relations between the proton and neutron TLFFs, 
although we will frequently compare hyperon and nucleon data. 

We do not make fits, however, when it will be the case,  straight lines will be used to highlight particular behaviors,  mainly to avoid too crowded 
graphs,. 
For instance, in semi-log plots in the $P$ variable, 
straight lines would correspond to exponential trends of the kind $e^{- aP}$, $a$ is a constant.  If one plots in the same graph all the available data on baryon TLFFS as a function 
of $q^2$ or $P$, the graph would be poorly readable. Thus, we will present only small 
samples of data within selected ranges, e.g., 
the near-threshold and the pre-asymptotic regions. 

%%%%%%%%%%%%%%%%%%%%%%%%%%%%%%%%%%%%%%%%%%%%%%%%%%%%%%
\section{Method}
\subsection{Analysis in the variables $P$ and $q^2$}
%%%%%%%%%%%%%%%%%%%%%%%%%%%%%%%%%%%%%%%%%%%%%%%%%%%%%%

\begin{figure} 
\begin{center} \includegraphics[width=8.5cm]{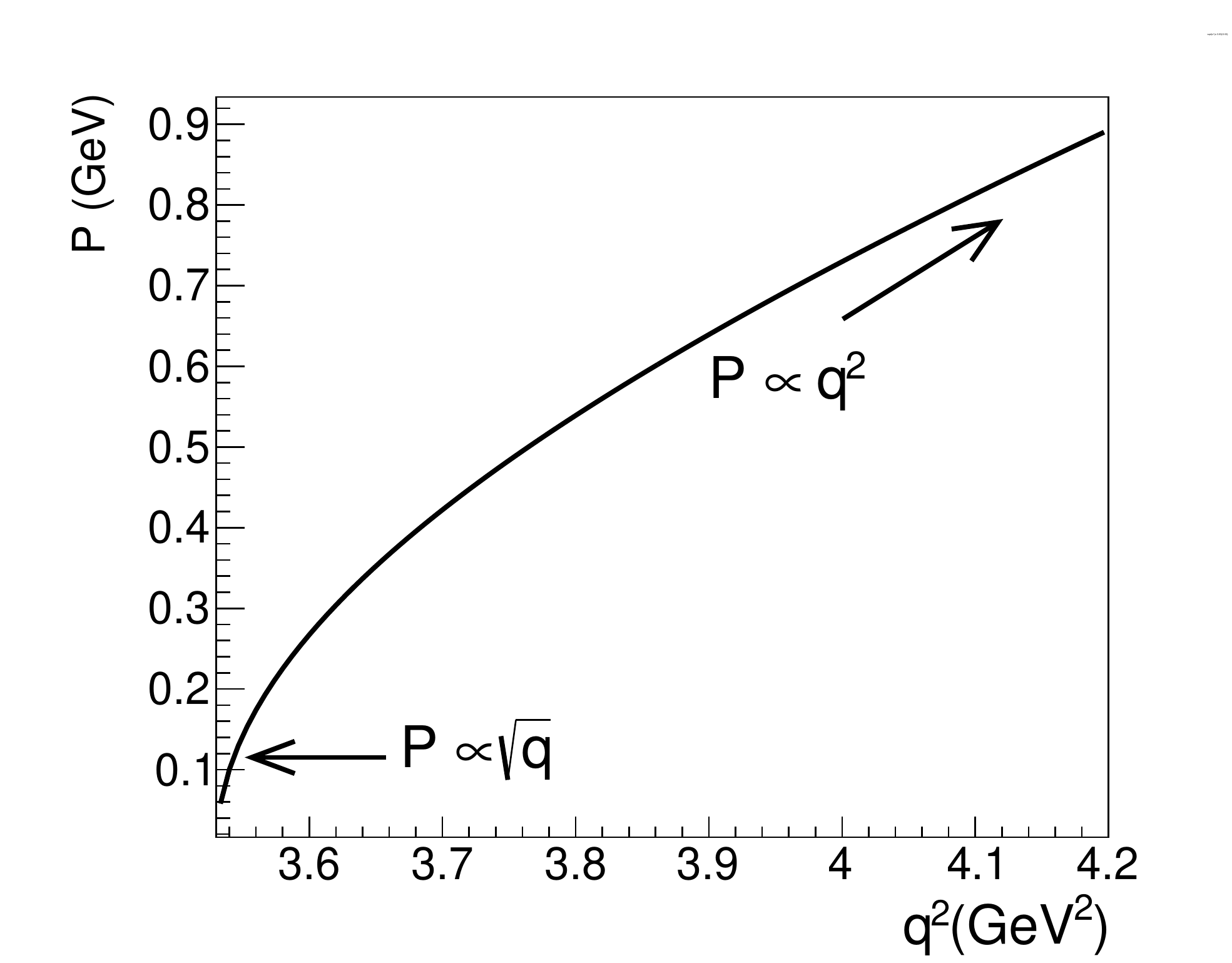}
\caption{Relation between $P$ and $q^2$ from 
Eq. (\ref{eq:PQ}), for the  $n\bar{n}$ channel. The red arrows and the formulae 
highlight the high-$q^2$ and the near-threshold limits, given by Eq.~\eqref{eq:PQ2}.} 
 \label{Fig:plab}
\end{center} 
\end{figure}

In this work, we apply a method introduced in Ref.
\cite{Bianconi:2015owa}, that consists using the variable
$P=|\vec P|$, i.e., the modulus of the baryon-antibaryon relative three-momentum,
in the baryon rest frame: 
\be 
P=\sqrt{\left( \frac{q^2}{4M_B^2}\ -\ 1 \right) q^2},
\label{eq:PQ} 
\ee
where $M_B$ is the baryon mass. In the near-threshold and the asymptotic regions 
such a variable behaves as 
\be
P \mathop{\simeq}_{q^2\to\infty} \frac{q^2}{2M_B}\,, \hspace{10mm} 
P \mathop{\simeq}_{q^2\to 4M_B^2}
\sqrt{4M_B\left(\sqrt{q^2}-2M_B\right)} \,, \label{eq:PQ2}
\ee
(we also examine data for the $\Lambda\bar{\Sigma^0}$ transition TLFF, where
the above equations are slightly modified due to the difference between the 
two hadron masses). As an example, for the neutron-antineutron
case, $P$ as a function of $q^2$ is shown in Fig.  \ref{Fig:plab}.

In the following we will compare TLFF data both in the  same $q^2$ and in the same $P$ conditions. As explicitly shown in Eq.~\eqref{eq:PQ}, $P$ is a function of $q^2$ 
and of the baryon mass $M_B$. It follows that, data at the same $q^2$ but for different channels,  i.e., different baryon masses $M_B$, have different values of $P$. In each channel, 
by definition $P=0$ when $q^2=4M_B^2$. For example, $P=0$ in the $\Lambda\bar{\Lambda}$ 
channel corresponds to a rather large value of $P$ in the nucleon channels, 
being $M_\Lambda > M_p$, $M_n$. 

If some common phenomenon characterizes the $n\bar{n}$ and the
$\Lambda \bar{\Lambda}$ final states produced through the annihilation reaction  $e^+e^-$
$\rightarrow$ $B\bar{B}$, with $B=n$ or $\Lambda$, then this would manifest itself  when comparing the outcome of the two channels at the same $q^2$, $ i.e.,$  at different values of $P$. Only in the proton versus neutron case  same $q^2$ and same $P$
are equivalent concepts, because of their similar masses. 

How  should be interpreted a phenomenon occurring  at the same $P$ in two channels with 
baryons having different masses? Since same $P$
implies different $q^2$,  this phenomenon is not an effect of any
communication between the channels, but it must be due to something that
independently but similarly takes place in both channels at a given
relative energy above threshold.

Another point distinguishes phenomena that appear as systematic 
in the  same $P$ or in the  same $q^2$ view. 
Assume that the $e^+e^-\to B\bar{B}$ reaction to proceed through a perturbative QCD (PQCD)
quark-gluon stage first, followed by the $B \bar{B}$ formation stage. The $B\bar B$ 
relative three-momentum $\vec P$ has no meaning in the initial stage. In fact, it cannot 
be defined until separated baryon and antibaryon emerge from the PQCD stage. So, whenever 
a phenomenon appears as systematic with respect to $P$ but not to $q^2$, 
it has to be associated with interactions between the forming or formed baryons 
in a stage where their physical separation takes place. 

%%%%%%%%%%%%%%%%%%%%%%%%%%%%%%%%%%%%%%%%%%%%%%%%%%%%%%%%%%%%%%%
\subsection{Interpolation of reference for proton form factor}
%%%%%%%%%%%%%%%%%%%%%%%%%%%%%%%%%%%%%%%%%%%%%%%%%%%%%%%%%%%%%%%

\begin{figure} 
\begin{center} \includegraphics[width=8.5cm]{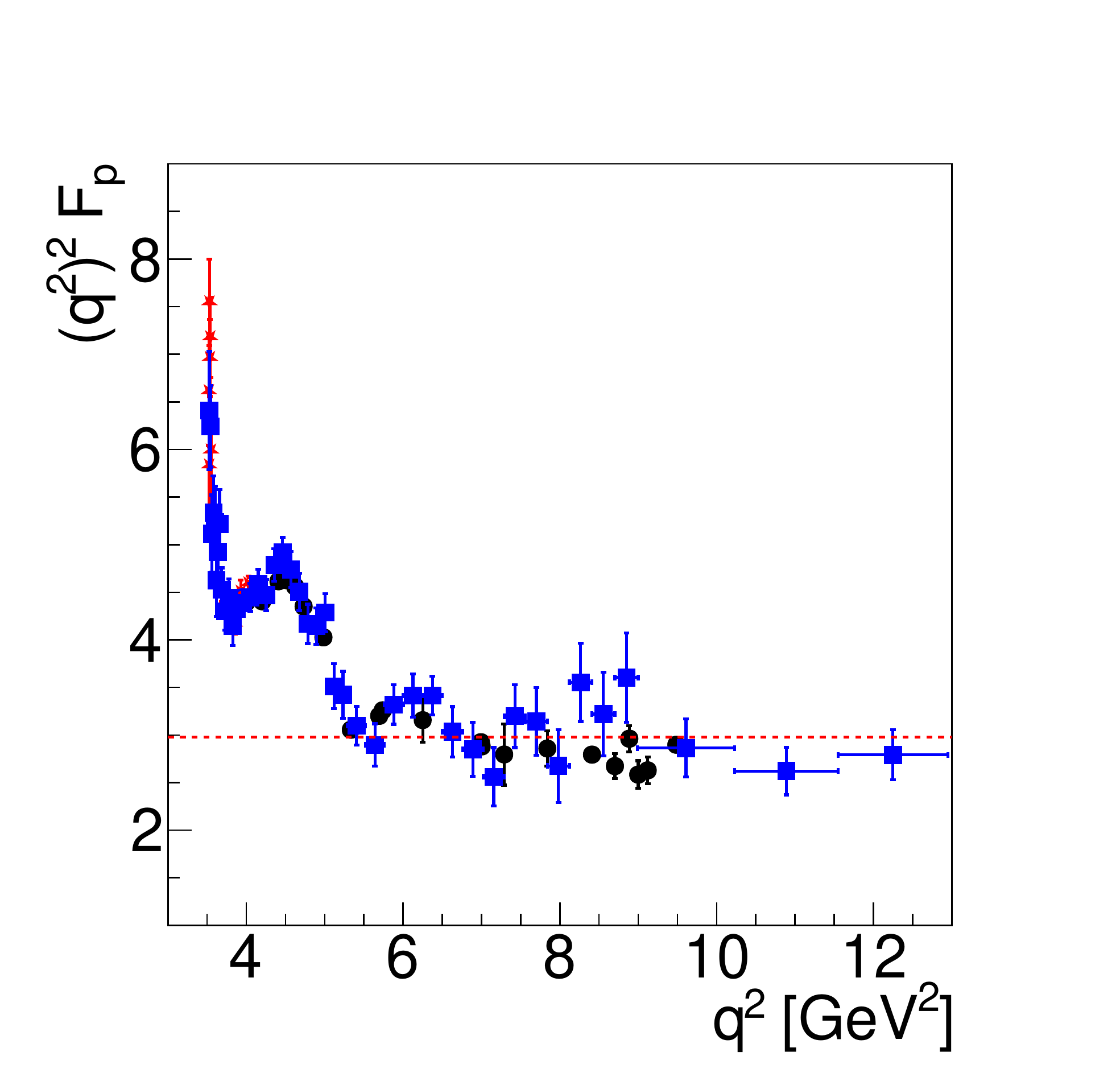}
  \caption{Simple interpolation of proton data in a preasymptotic region. 
    The data are $F_p$ data from the BABAR Collaboration
\cite{Lees:2013xe,Lees:2013uta} (blue squares),  the CMD-3
Collaboration  \cite{Akhmetshin:2015ifg, CMD-3:2018kql} (red stars), and the  BESIII Collaboration  \cite{BESIII:2019hdp} (black circles). The dashed red horizontal line represents an interpolation of proton data for
5 GeV$^2$ $<$ $q^2$ $<$ 13 GeV$^2$. This is not a precise fit but is useful for later comparison with baryon data presenting special features in that region. } 
\label{Fig:protonQ2} 
\end{center} 
\end{figure}

\begin{figure} 
\begin{center} 
\includegraphics[width=8.5cm]{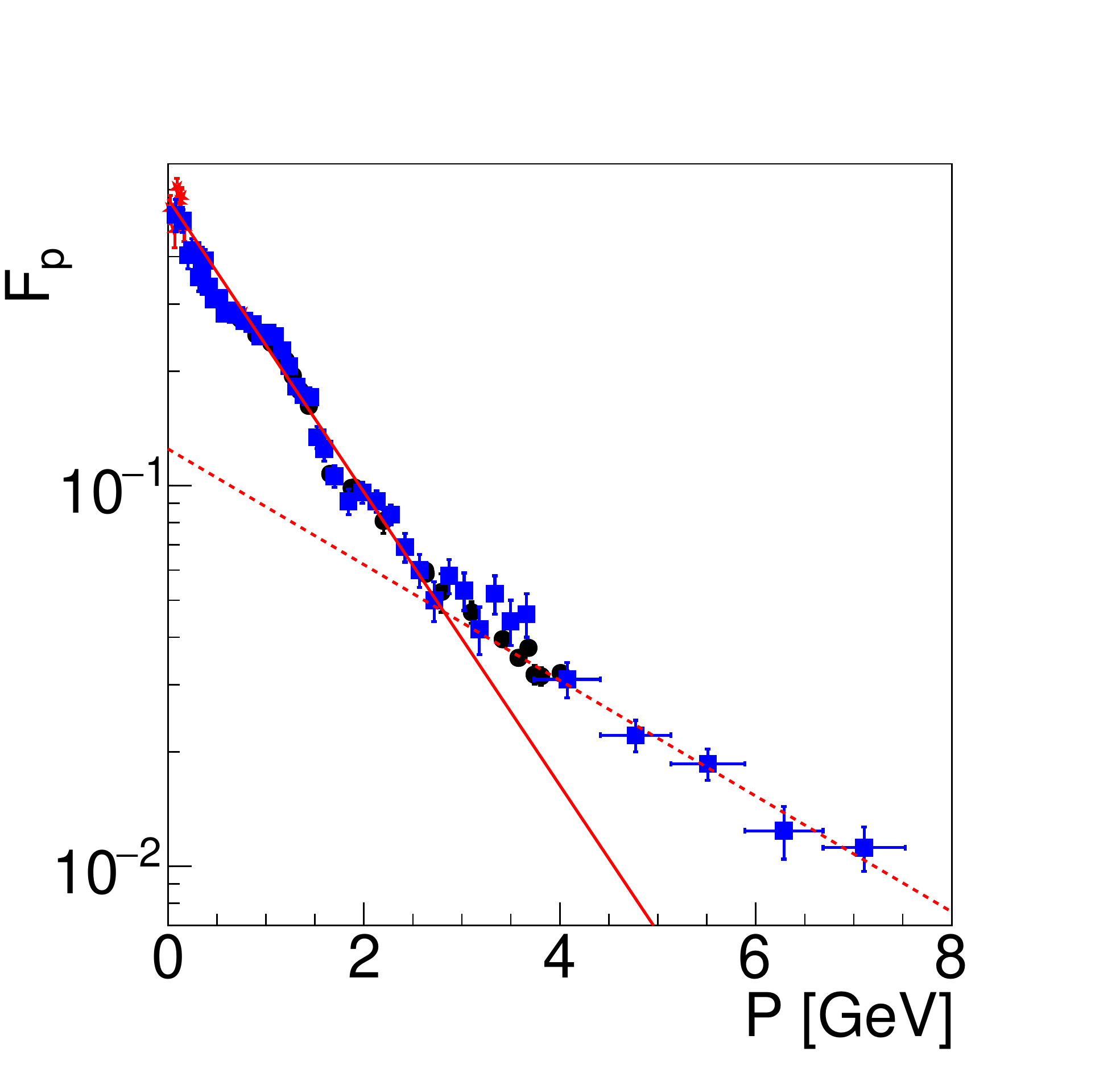}
\caption{Near-threshold and asymptotic interpolations of proton data
  versus $P$. The data are as in Fig. \ref{Fig:protonQ2}.  The two straight lines show interpolations of the proton data for $P$
smaller(larger) than 2.75 GeV, red solid line (red dashed line).
Because of the semilog plots, these correspond to $e^{- a P}$
shapes. Although we do not attribute any special meaning to these
interpolations, they are used as a reference in later figures showing
other baryon data, to compare them with proton data trends at small and large $P$. }
\label{Fig:protonPL} 
\end{center} 
\end{figure}

We will often use proton data as a reference, because of their unrivaled
quality in terms of precision, range, continuity, and coherence 
between different measurements. However, in order not to overload the figures with data points, instead
of the proton data, we may use the straight lines shown in Figs.
\ref{Fig:protonQ2} and \ref{Fig:protonPL} that indicate the quantities
described in the following: 

a) The ``asymptotic'' horizontal linear fit of $(q^2)^2 F_p(q)$ in the
$q^2$ range 5.5-13 GeV$^2$. We expect  $(q^2)^2 F_p(q)$ $\approx$
constant at large $q^2$, and this is the reason  behind the above $fit$.
Actually, two BABAR points with large error bars at $q^2$ between 20 and
25 GeV$^2$ (not shown in this figure) lie slightly below this line,
suggesting that a really asymptotic trend has not been reached. For 5
GeV$^2< q^2<  13$  GeV$^2$, the above interpolation well averages the
proton data and it is therefore later used as a reference in the figures
showing baryon data versus $q^2$.

b) Two straight-line fits of $F_p$ as a function of  $P$ in semilog plot. These are
fits of the form $e^{-aP}$. The former fit is a basic ``background''
fit (as in Eq. (\ref{eq:osc0})) interpolating the oscillating proton
data in the near-threshold region, up to about $P=3$ GeV. The latter
well reproduces all the proton data in the region $P >$ 4 GeV  where
oscillations are not visible anymore, and can be considered
$asymptotic$, since it involves the largest $q^2$ available data.

%%%%%%%%%%%%%%%%%%%%%
\section{Results}
\subsection{Near threshold $P\le 2$ GeV range}
\begin{figure} 
\begin{center} 
\includegraphics[width=8.5cm]{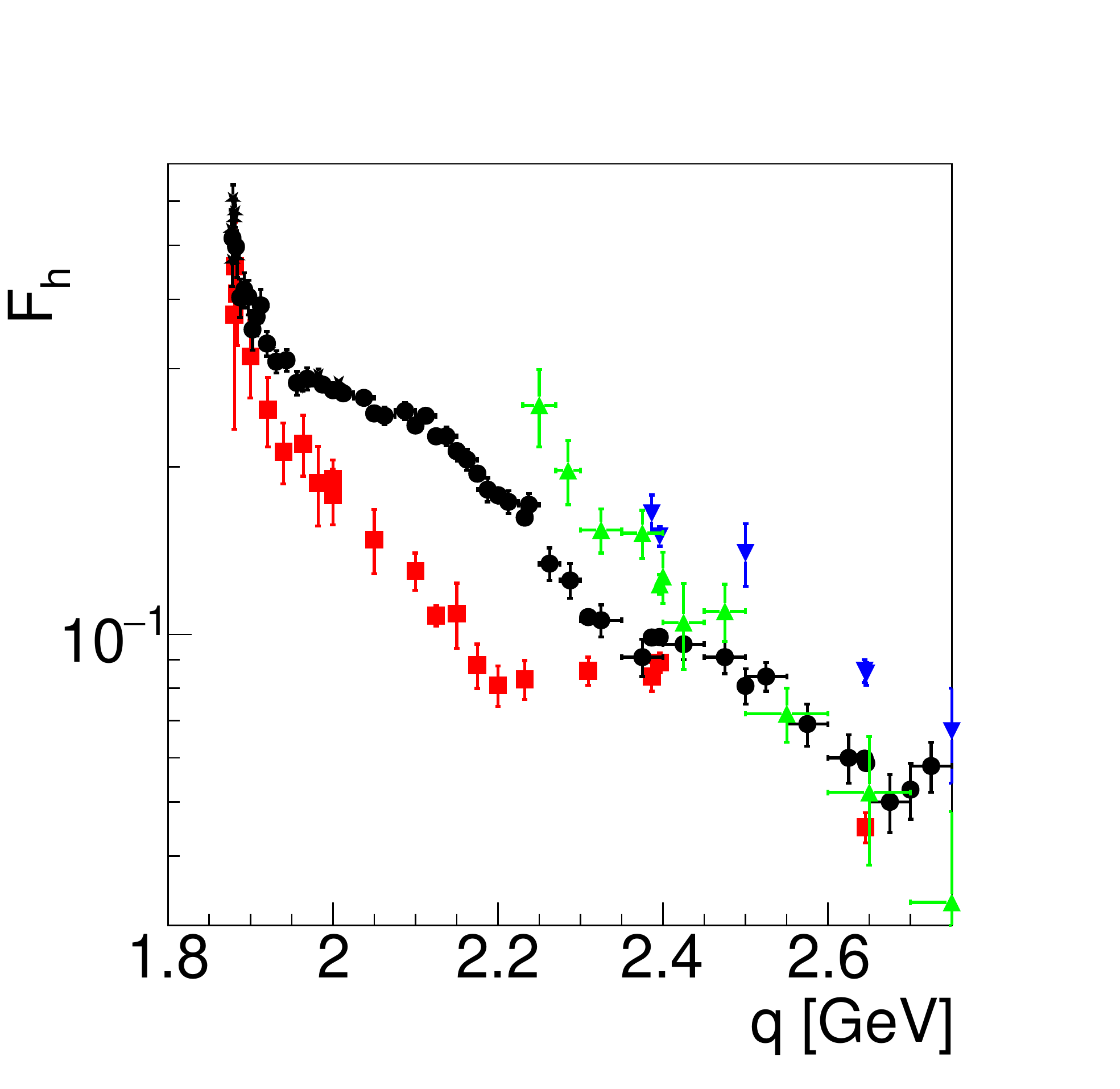}
\caption{ TLFFs of proton (black circles), neutron (red squares),
$\Lambda$ (green triangles) and $\Sigma^+$ (blue triangledowns) versus
$q=\sqrt{q^2}$ in the region or relatively small $q$. Two charged and
two neutral species characterized by small error bars have been selected
for this figure. More species will be considered in the following, but
presenting them altogether would make the picture difficult to read. }
\label{Fig:C11} 
\end{center} 
\end{figure}

%%%%%%%%%%%%%%%%%%%%%%%%%%%%
The near-threshold behavior of the four lowest-mass species, two neutral
(neutron and $\Lambda$) and two charged (proton and $\Sigma^+$) is shown
in Fig.  \ref{Fig:C11} as a function of $q=\sqrt{q^2}$. Although they show a similar qualitative behavior, it is difficult to find shared trends at a
precise level, even if one substitutes proton and neutron with their
$background$ components, i.e., the average over local oscillations.
The first difficulty is given by the different thresholds.

\begin{figure} \begin{center}
\includegraphics[width=8.5cm]{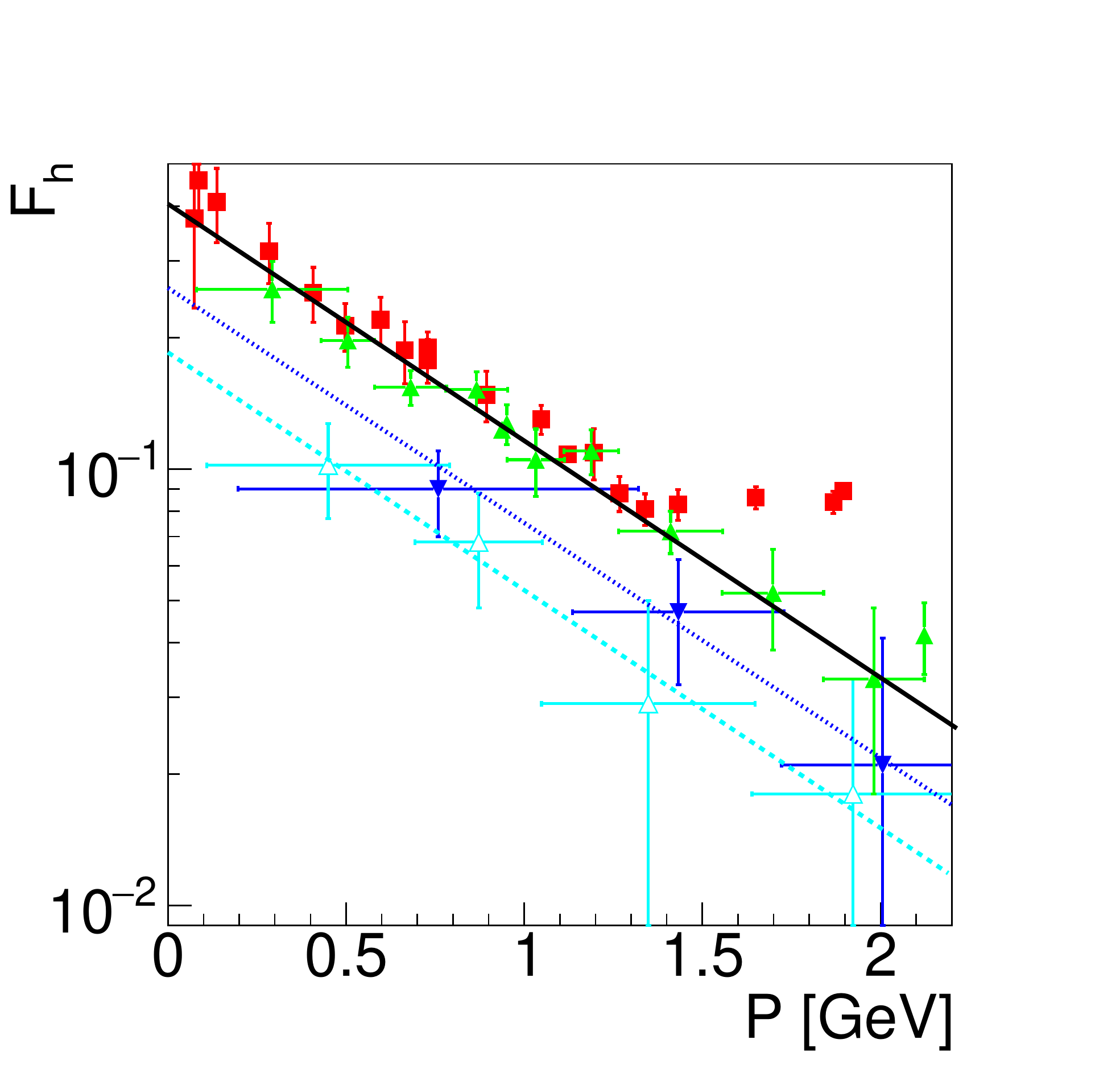} 
\caption{ TLFFs of neutron (red
squares), $\Lambda$ (green triangles),  $\Sigma^0$ (blue triangledowns),
and $\Lambda\bar\Sigma^0$ (open cyan triangles) transition TLFF (a neutral
$\Lambda\bar\Sigma^0$ pair is produced), as functions of $P$ in the
near-threshold region. A solid black straight line is used to
interpolate the neutron and $\Lambda$ points, a dotted blue line  with
the same slope the $\Sigma^0$ points, and a dashed cyan line the
$\Lambda\bar\Sigma^0$ points. Because of the semilog plot, these straight
lines correspond to $A e^{-a P}$ exponential functions with
different normalizations $A$ but same slope $a$. }
\label{Fig:neutralPL} 
\end{center} 
\end{figure}

Neutral baryon TLFFs are shown in Fig.  \ref{Fig:neutralPL} as a
function of $P$. We notice that in a semilog plot all the points
organize themselves along straight lines with the $same$ $slope$.  
The exception to this kind of neutral-baryon rule is  
near-threshold $\Xi^0$ points, not reported in this figure, rise instead of falling.

Neutron data practically coincide to those of the $\Lambda$ up to $P=1.2$ GeV, then they change slope, anticipating what the other TLFFs do at larger $P$. 
A straight line allows to appreciate the slope. Two more straight lines parallel 
to the neutron and $\Lambda$ line are shown. One is compatible with the $\Sigma^0$ points, 
the other one with the data of the $\Lambda \bar\Sigma^0$ transition TLFF.

\begin{figure}
\begin{center} 
\includegraphics[width=8.5cm]{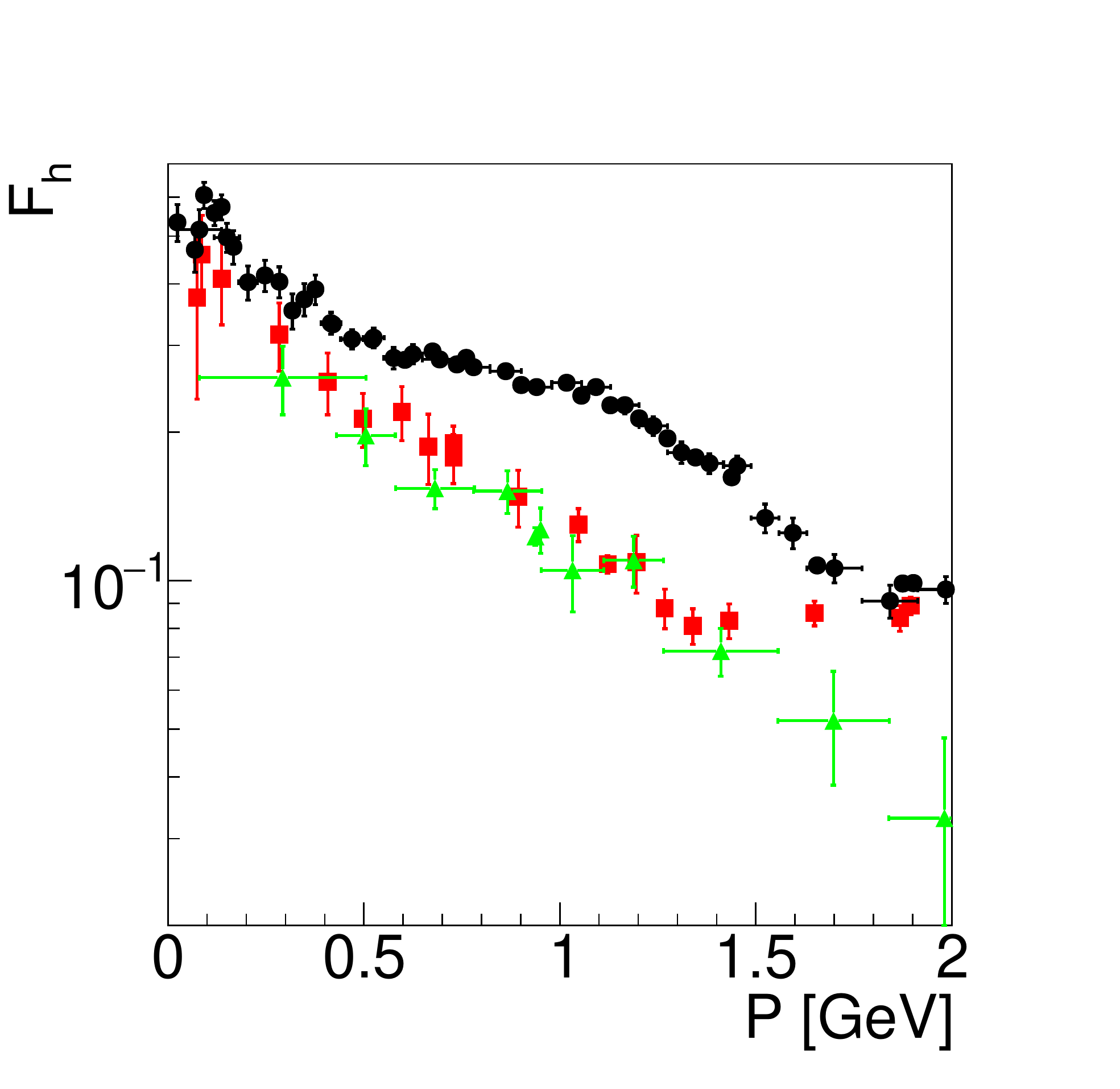} 
\caption{
Compared TLFF of neutron (red squares), $\Lambda$ (green triangles) and
proton (black circles) versus $P$. One may notice the separation of proton points from neutron and $\Lambda$ just above threshold.
Neutron and $\Lambda$ points
follow the same trend up to $P=1.5$ GeV. Above this value, neutron data 
converge to the proton data. } 
\label{Fig:reduced2} 
\end{center}
\end{figure}

Neutron, $\Lambda$, $\Sigma^0$ and $\Lambda \bar\Sigma^0$ 
TLFFs have a very similar near-threshold behavior, which, instead, is quite different 
from that of the other baryons. In Fig.~\ref{Fig:reduced2} the neutron and $\Lambda$ 
TLFFs are compared to the proton one in the same range. 
A plain observation shows two facts: 
\begin{itemize}
\item[(a)] Near the threshold, all data coincide, but just above  
	a separation takes place with a softer average slope for the proton. 
\item[(b)] At $P=1.2$ GeV the neutron values move away from the $\Lambda$ ones and reach the proton data.
\end{itemize}
This relative proton-neutron behavior is the ground for the joint
oscillation fit with a phase difference proposed in Ref.
\cite{BESIII:2021tbq}. Any phase difference requires the two to
immediately separate and periodically rejoin. Whatever the
interpretation, the initial average slope of the proton $F_p$ is
different from the common slope of neutron and $\Lambda$, that is the
same as for $\Sigma^0$ and $\Lambda \bar\Sigma^0$. 

\begin{figure} 
\begin{center} 
\includegraphics[width=8.5cm]{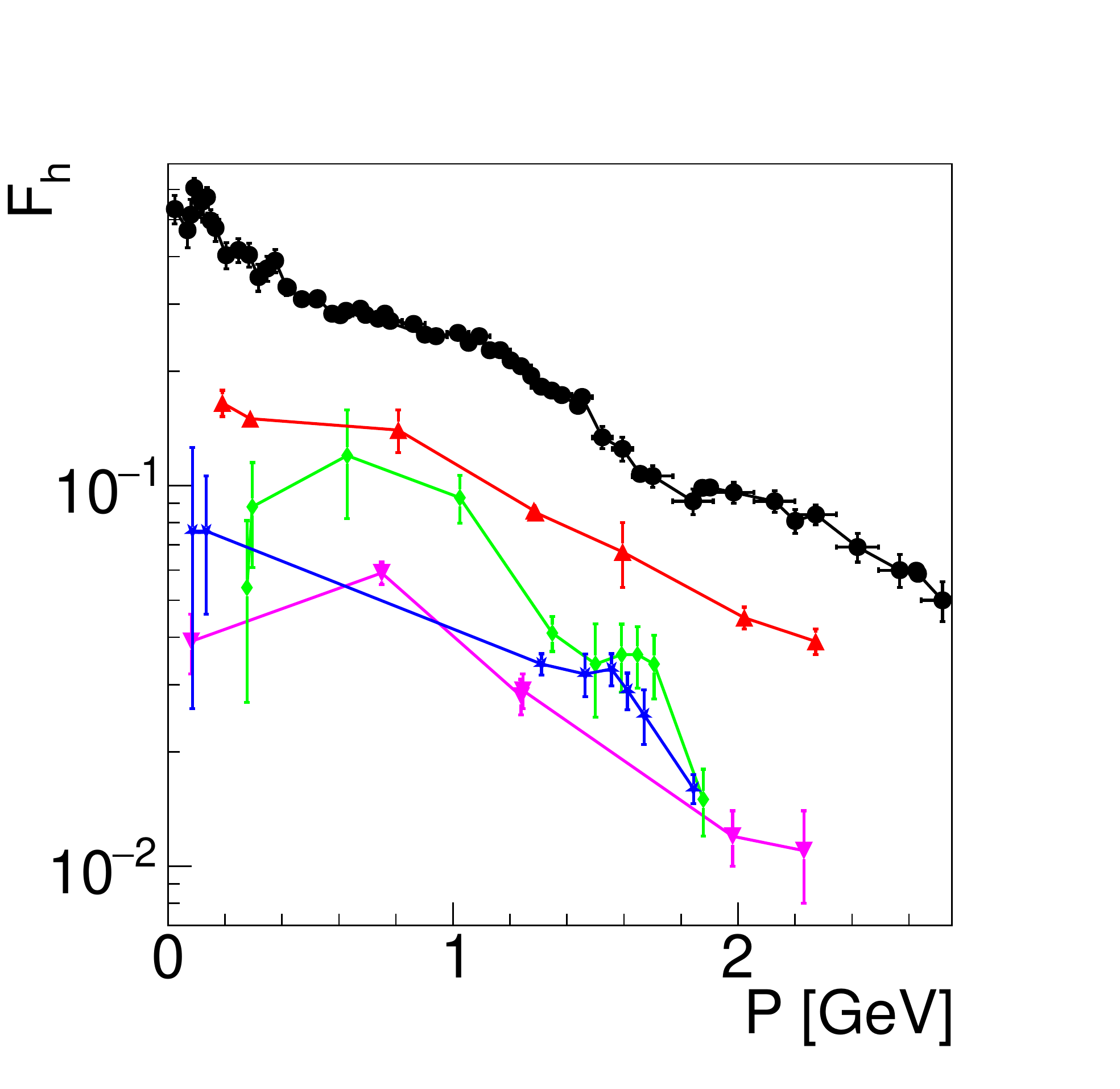}
\caption{ $F_{h} $ versus  $P$ or $\Sigma^+$ (red triangles), $\Sigma^-$
(magenta triangledowns), $\Xi^-$ (blue stars),  and $\Xi^0$ (green
  diamonds), together with proton data (black circles). Since points of different sets overlap, we have used
  segments to join the points of the same set (in the proton case, we have three sets of data, see fig.\ref{Fig:protonQ2}). 
  The most remarkable feature here is the presence of a
correlated peak/shoulder for $\Sigma^+$ and $\Sigma^-$, and the presence
of $two$ peaks in the case of $\Xi^0$, one correlated with $\Xi^-$ and
one with $\Sigma^-$ and $\Sigma^+$. These data have a correspondence
with the peak-rich structure of the proton (black solid circles). }
\label{Fig:chargedPL} 
\end{center} 
\end{figure}

\begin{figure} 
\begin{center}
\includegraphics[width=8.5cm]{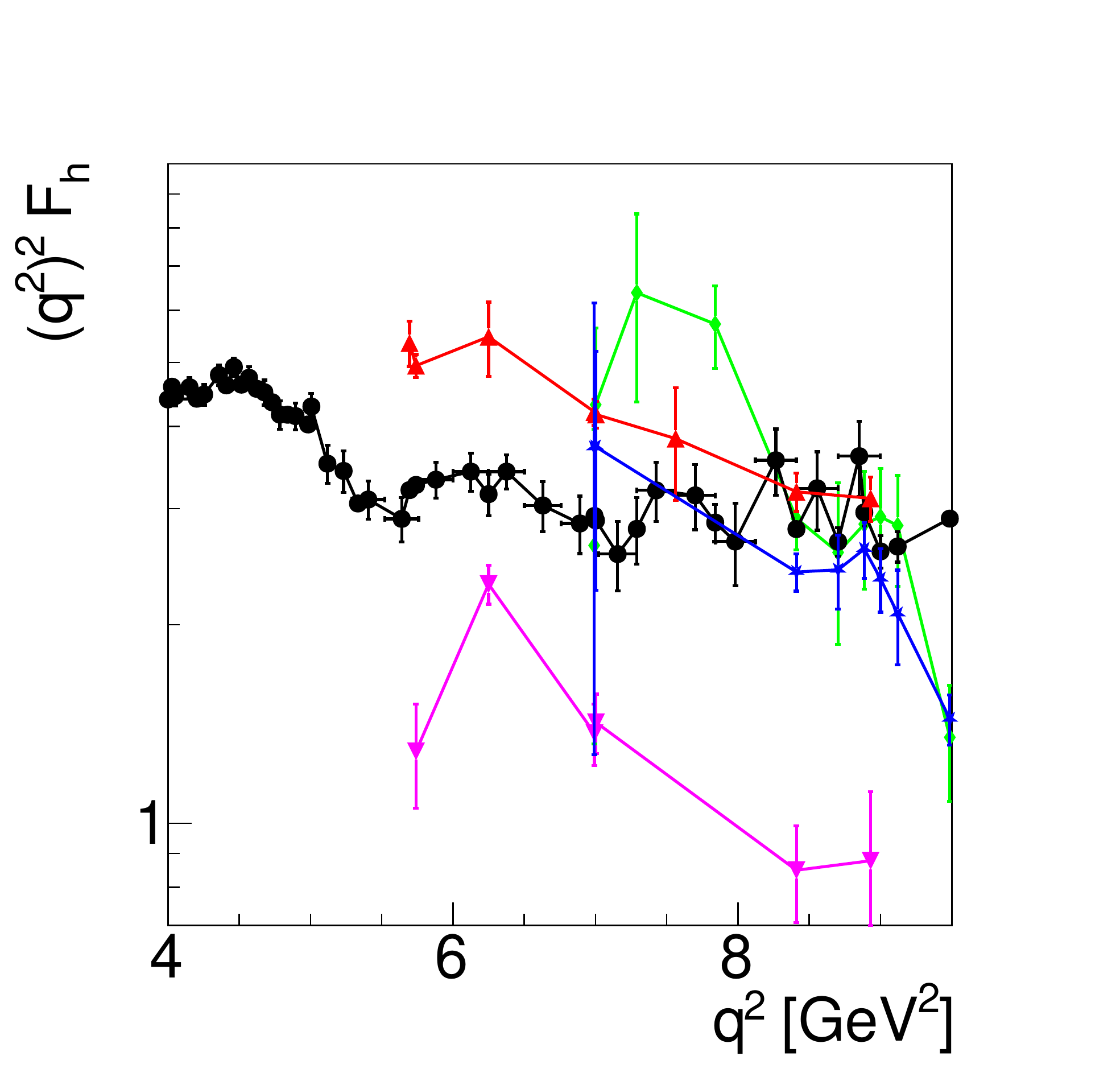} 
\caption{
$(q^2)^2 F_h $  for $\Sigma^+$ (red triangles), $\Sigma^-$
(magenta triangledowns), $\Xi^-$ (blue stars),  and $\Xi^0$ (green
  diamonds), together with proton data (black circles).
Since points of different sets overlap, lines connect the sequence of points of the same set (in the proton case, three sets of data are drawn, see Fig. \ref{Fig:protonQ2}).The  strong correlation between the peaks of different data is evident and suggests
that the fourth proton maximum is not a statistical fluctuation, but
consists of at least one and perhaps a pair of maxima, correlated with a
peak-dip-peak sequence part of which has a correspondence, for example, in the
$\Xi^0$ data.} 
\label{Fig:chargedQ} 
\end{center} 
\end{figure}

The charged FFs presently available are compared with the proton one,
together with the $\Xi^0$ isospin partner of $\Xi^-$  to check for
correlations in the variable $P$ in Fig.~\ref{Fig:chargedPL} and in the variable $q^2$ 
in Fig.~\ref{Fig:chargedQ}. 

These data, plotted as a function of $P$, show that the behavior of the charged
baryons and  of $\Xi^0$ is different with respect to  the considered
neutral ones but presents common peculiarities. In particular a rich
peak/shoulder structure is present.

Plotting for the same baryons $(q^2)^2 F_{h}(q^2) $ with respect to  $q^2$
shows a good correlation between the oscillations of the baryons and the
second, the third and fourth oscillation that could actually consist of
two separate peaks. Indeed, although the proton points appear as
randomly oscillating, their oscillations are well correlated with
the oscillations of all the other baryons.

In the $P$-plot we find a coincidence between
peak/shoulders of $\Xi^0$ and $\Sigma^\pm$, but a similar coincidence
occurs between $different$ maxima in the case of the $q^2$-plot.  This looks like a
random effect due to the mass difference between the two channels
coinciding with the shift between different maxima in the same channel.
Summarizing what we feel we may conclude from these two figures: 

\begin{enumerate}
	
\item Maxima do evidently exist in the charged baryon data and their
correlation shows that they are not due to data fluctuations over
underestimated statistical errors. In the case of $\Xi^0$ at least two
maxima are well visible and both coincide with maxima of other baryons.

\item The fourth proton oscillation maximum may not be a random fluctuation of
the data in a region where error bars have the same magnitude of the
oscillation amplitude.

\item The well correlated oscillations of $\Sigma^+$ and $\Sigma^-$ effective TLFFs are
less predictable than one could expect, since these two baryons are not
antiparticles of each other and have a slightly different mass. Their
third isospin partner $\Sigma^0$  behaves in a different way, as shown
in other figures. 

\end{enumerate}

%%%%%%%%%%%%%%%%%%%%%%%%%%%%%%%%%%%%%%%%%%%%%%%%%%%%%%
\subsection{Comparison among Strangeness 0 and 1, neutral versus proton,
in a broader range}
%%%%%%%%%%%%%%%%%%%%%%%%%%%%%%%%%%%%%%%%%%%%ù
Proton data cover a much larger range than any other. As shown in Refs.
\cite{Bianconi:2015owa,Bianconi:2016bss} where a 5-10 \% oscillating
modulation is subtracted,  the leading background follows a regular trend, 
described by a $q^2$ tri-pole. 
However, the TLFFs of neutral baryons shown here present a steeper near-threshold fall. 

This argument does not change comparing data at the same $q^2$. Measured values of the effective 
TLFF of all baryons near their thresholds are larger than the proton one,
but they decrease within shorter energy intervals, i.e., they have a steeper decreasing behavior. 

On the other side the arguments leading to the $(1/q^2)^2$ asymptotic behavior with logarithmic corrections for the leading $F_1$ helicity-conserving component of the TLFFs, have not met serious challenges up to now. So we face three possibilities.
\begin{itemize}
\item[{\bf A.}] Neutral baryon  TLFFs follow an asymptotic behavior as $\simeq A/(q^2)^2$ while  $F_p \simeq B/(q^2)^2$ with $A\ll B$.

\item[{\bf B.}] In the region separating near-threshold and asymptotic
$q^2$, one strong oscillation at least is present, that allows the
neutral FFs to rise up to proton-like values at large $q^2$.

\item[{\bf C.}] Because of a suppression of the leading
helicity-conserving $F_1$ term, the $1/(q^2)^2$ trend is not reached in the investigated $q^2$ region. 
\end{itemize}

The quark counting rule predicting the $\simeq A/(q^2)^2$ asymptotic behavior 
does not constrain the magnitude of $A$. On the other side, for $A$ to be very different 
from case to case, one should imagine either a suppression mechanism, acting in some cases 
but not in general, or a very different formation path for different  baryon-antibaryon pairs. 
At large $q^2$, SU(3) symmetry should be restored, implying differences like $\sqrt{3}/2$ 
or so between the different baryon FFs. These considerations would suggest that the case 
{\bf B} of the previous list should be the most likely, what  will be checked in future data. 

For the effective FFs of the charged strange baryons and of the neutral $\Xi^0$, only the 
near-threshold data shown in the previous figures exist, so they are not considered 
in this section. For the neutron and the strange baryons effective TLFFs 
the data span a medium kinematical range, meaning the region of $P$ up to 10 GeV. 
Their behavior is however not homogeneous, so they are discussed separately,  using proton data as a reference. 
Specifically for $\Lambda$, recent measurements are available for  $q^2$ values larger than those of the other neutral baryons. 

\begin{figure} 
\begin{center} 
\includegraphics[width=8.5cm]{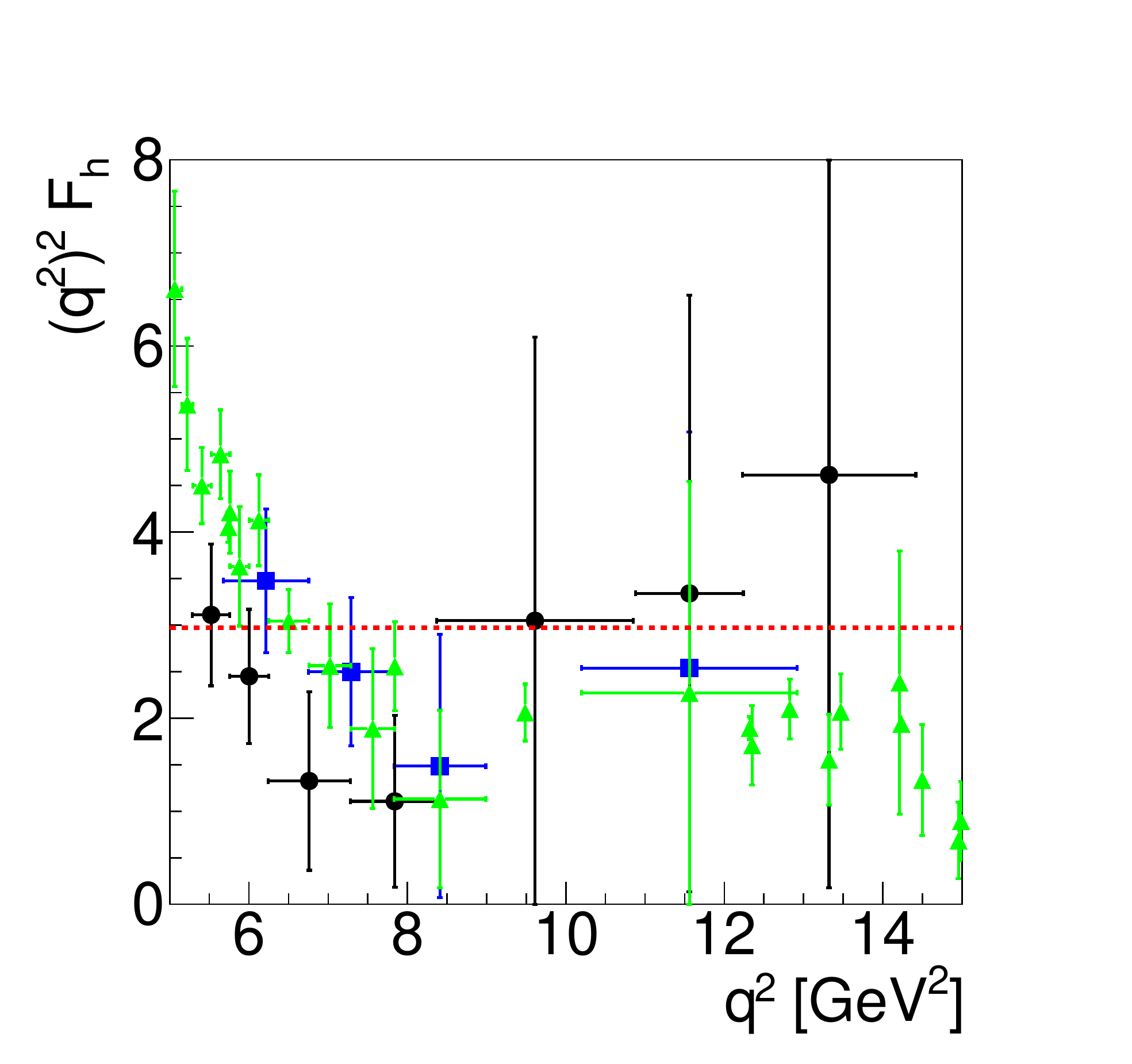}
\caption{$(q^2)^2 F_h $ as a function of  $q^2$ for $\Lambda$
(green filled triangles), $\Sigma^0$ (blue squares) and transition
$\Lambda\bar\Sigma^0$ (black circles) data.  The horizontal red dashed
line is the proton $preasymptotic$  fit of Fig. \ref{Fig:protonQ2}. A
coherent minimum-maximum sequence is visible in the cumulative set  of
strange neutral baryons, that does not show simple
correlations with the structures in Fig.  \ref{Fig:chargedQ}. Roughly,
the minimum here is not far from a maximum in Fig.  \ref{Fig:chargedQ}.
} 
\label{Fig:A11} 
\end{center} 
\end{figure}

Figure~\ref{Fig:A11} shows a clean example of a case {\bf B}. In the region $\sqrt{q^2}= 5-15$ GeV, 
the quantity $(q^2)^2 F_h(q^2)$ is almost constant for the proton, 
it is represented by a thick horizontal dashed line. On the contrary, 
for the TLFF of $\Lambda$ and $\Sigma^0$, and the $\Lambda\bar\Sigma^0$ 
transition TLFF an initial steep decrease 
is followed by a rise that leads the data at the proton values again. 

%%%%%%%%%%%%%%%%%%%%%%%%%%%%%%%%%%%%%%%%%%%%%%%%%%
\subsection{Effective form factors of the $\Lambda$ compared to the nucleons}
%%%%%%%%%%%%%%%%%%%%%%%%%%%%%%%%%%%%%%%%%%%%%%%%%%

\begin{figure} 
\begin{center}
\includegraphics[width=8.5cm]{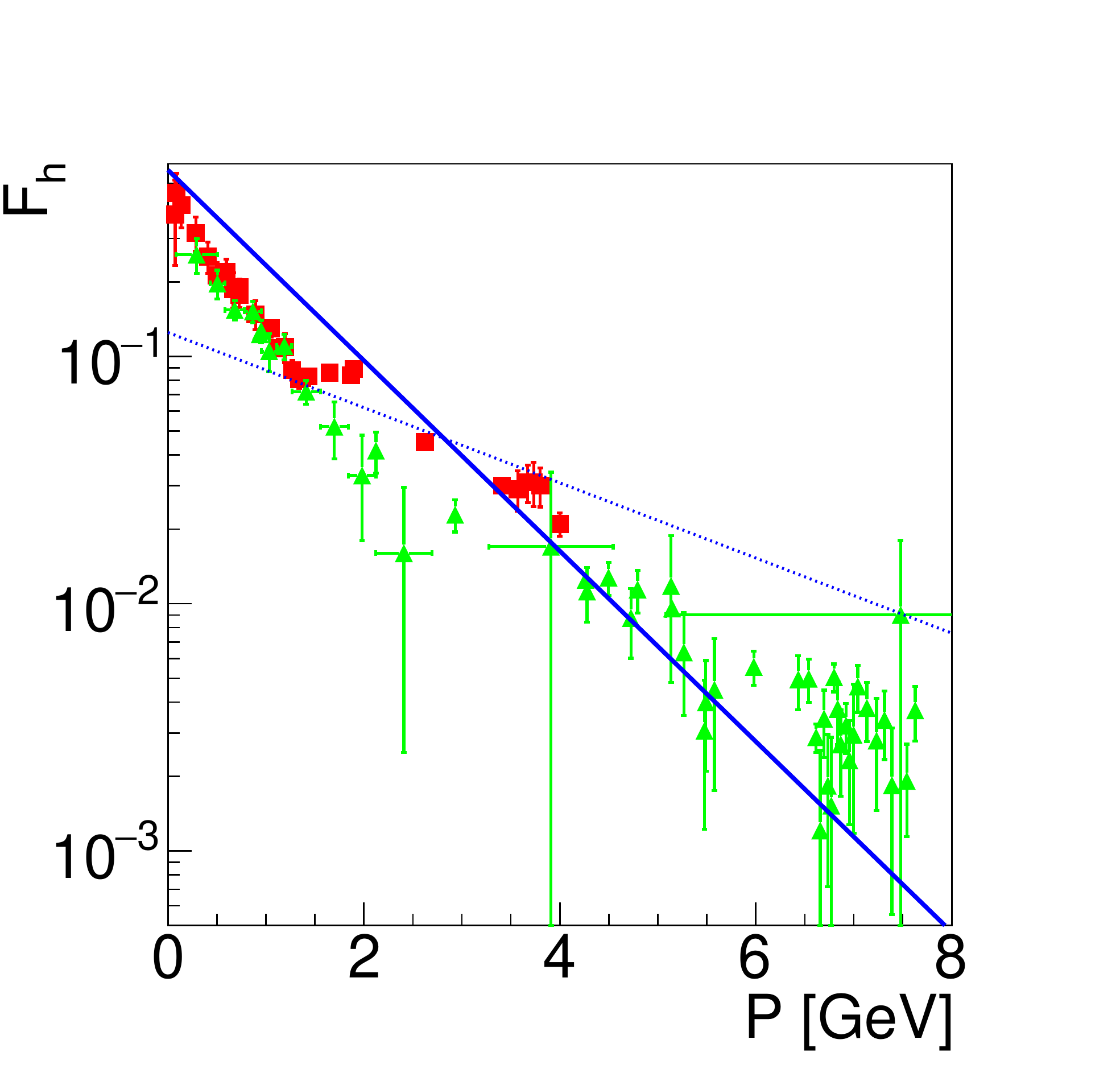} 
\caption{$\Lambda$
and neutron TLFFs as a function of $P$.  The straight lines are the proton TLFF 
interpolations of Fig.  \ref{Fig:protonPL} at small-$P$ (solid blue line) 
and large-$P$ (dotted blue line). At small $P$ neutron and
$\Lambda$ data coincide and have a steeper fall than proton data. At
$P> 2$  GeV,  neutron data reach the proton values. Above $P=$ 4 GeV no
neutron data are available, while $\Lambda$ data are in the average
steeper than proton data,  and show hints of minima  at
5.5, 6.7, 7, 7.5 GeV.  In this region only proton and $\Lambda$ data
are available.} 
\label{Fig:asym3} 
\end{center} 
\end{figure}

Figure~\ref{Fig:asym3} shows the data of the $\Lambda$ 
and the neutron effective TLFFs, together with the straight lines 
that interpolate the behavior of proton data at $P$ below and above 4 GeV 
reported in Fig.~\ref{Fig:protonPL}. Both neutron and $\Lambda$ data, 
after the initial steep decrease, show a rise as a tendency to reach the proton values. 

Such a tendency is fully realized in the case of the neutron at $P \approx 2$ GeV and 
$P \approx 4$ GeV. What happens precisely in between is not clear yet as  there 
are no data at larger energies. 

For $P>5 $ GeV, neutron data are not present anymore and only $F_\Lambda$ 
may be compared with $F_p$. The surprising recent BESIII $\Lambda$ data at large 
$P$ follow the extrapolation of the small-$P$ proton data. 
Visibly, the $F_\Lambda$ values at large $P$ spread over a region where 
$F_\Lambda$ is smaller than $F_p$ by a factor that ranges from 0.1 up to 0.6. 

The data also seem to suggest that $F_\Lambda$ may present some deep minima in this region, 
however the data points are dense and it is difficult to distinguish physical from statistical fluctuations. 

%%%%%%%%%%%%%%%%%%%%%%%%%%%
\section{Conclusions}
%%%%%%%%%%%%%%%%%%%%%%%%%%

The existing data on baryon TLFFs were collected and compared in the
attempt of finding common features and highlight the kinematical region
and the reactions where future data are highly desirable.
The comparison of the data in terms of two kinematical variables $P$ and $q^2$, allows to visualize 
possible correlations and make conjectures about their origin.  

The presence of three oscillations in proton TLFF data is a well known fact by now, but a visible fourth oscillation maximum
is still ambiguous because of errors.
The presence at the same $q^2$ of oscillation maxima of other baryon TLFFs (Fig. \ref{Fig:chargedQ})  
supports the physical reality of this maximum in the proton data.

The $F_\Lambda$ at high $q^2$ shows possible structures, with up to four minima that could be very deep, or just
be due to local error dilatation. Further precise data in this region 
may clarify these structures, as well as the asymptotic behavior, clearly steeper than the proton one. 

Neutron and $\Lambda$ data show several similarities, but continuous neutron data stop at $\sqrt{q^2}$ $=$ 2.4 GeV, with a few more scattered points at larger $q^2$. These may be 
compatible either with proton-like regular oscillations, or with
$\Lambda$-like behavior. Continuous neutron data over 2.4 GeV are needed to
clarify the what could be a complex network of correlations between neutron, proton and $\Lambda$. 

Baryon data near their respective threshold can be divided into two groups with very different behavior. This difference is so marked that more precise data could not overthrow it. 
The case of the strangeness-1 neutral baryons, including the transition $\Lambda\bar\Sigma^0$ TLFF, 
is especially interesting since they seem to present the same marked down-up-down change of slope at
$q^2$ $\approx$ 10 GeV$^2$, possibly overlapping. The neutron shows a very similar feature at smaller
$q^2$. So, more precise data in this region could show interesting and unpredicted phenomena.

Summarizing, although the proton data oscillation is already something that escapes obvious interpretations, the
landscape of the baryon data could be far richer, and new precise  measurements will surely show surprising features. 

Further information from TLFFs is contained in the time-reversed reactions induced by antiproton and antineutron beams. In next future, the PANDA
experiment at FAIR is expected to run an important program of FF
measurements~\cite{PANDA:2016fbp}, by considering the processes
 $p\bar p\to e^+e^-,\  \mu^+\mu^-$ ~\cite{PANDA:2020jkm,Dbeyssi:2011tv}.

Further discussion of the presented data, in connection with 
possible formation mechanisms, will be presented in a forthcoming paper.  

\section{Acknowledgments}

Thanks are due to Simone Pacetti for interesting discussions and a careful reading of the draft.

%\bibliography{Biblio}

\end{document}